\documentclass[twocolumn,prd,aps,showpacs,showkeys,amsmath,amssymb,nofootinbib]{revtex4-1}
\usepackage{bm}

\usepackage{color}
\usepackage{amsmath}
\usepackage{amsfonts}
\usepackage{verbatim}
\usepackage{amssymb}
\usepackage{graphicx}
\usepackage{epstopdf}
\usepackage{mathrsfs}
\usepackage{epsfig}
\usepackage{slashed}
\usepackage{bbold}%Matriz identidade
\usepackage{color} %Usado para colorir palavras.

\providecommand{\U}[1]{\protect\rule{.1in}{.1in}}
\newcommand{\bl}{\boldsymbol}

\newcommand{\ph}{\phantom}

\begin{document}

%A Petrov Type II Exact Vacuum Solution

\title{A Class of Higher-Dimensional Solutions of Einstein's Vacuum Equation  }
\author{Gabriel Luz Almeida and Carlos Batista}
\email[]{carlosbatistas@df.ufpe.br}
\affiliation{Departamento de F\'{\i}sica, Universidade Federal de Pernambuco,
Recife, Pernambuco  50740-560, Brazil}

%\date{\today}

\begin{abstract}
A new class of higher-dimensional exact solutions  of Einstein's vacuum equation is presented. These metrics are written in terms of the exponential of a symmetric matrix and when this matrix is diagonal the solution reduces to higher-dimensional generalizations of Kasner spacetime with a cosmological constant. On the other hand, the metrics attained when such matrix is non-diagonal have more intricate algebraic structures. Such solutions have not been presented in the literature yet.
\end{abstract}
\keywords{Exact solutions; Higher-dimensional general relativity; Kasner solution}
\maketitle

\section{Introduction}

The aim of this article is to fully integrate Einstein's vacuum equation, with a cosmological constant, for the class of $(n+2)-$dimensional spaces with the line element
\begin{equation}\label{LineElement1}
  ds^2 = (S_x + S_y)\left[ H^{ij}_x d\sigma_i d\sigma_j + \frac{dx^2}{\Delta_x^2} + \frac{dy^2}{\Delta_y^2}  \right] \,,
\end{equation}
where the indices $i,j$ run from $1$ to $n$. The subscripts $x$ and $y$ indicate that a function depends just on the coordinate $x$ and $y$ respectively. Thus, for instance, the components of the $n\times n$ symmetric matrix $H^{ij}_x$ are functions of $x$,  $H^{ij}_x = H^{ij}_x(x)$. In particular, none of these functions depend on the coordinates $\sigma_i$, so that these spaces are endowed with $n$ commuting vector fields, namely $\partial_{\sigma_i}$. Note that the functions $\Delta_x$ and $\Delta_y$ can be easily gauged away by redefining the coordinates $x$ and $y$. Nevertheless, instead of setting these functions to 1, we shall keep them and make a more convenient choice afterwards. As we shall see in the sequel, the general solution of Einstein's vacuum equation can be elegantly written in terms of the exponential of a symmetric constant matrix $\bl{Q_0}$. In particular, the special case in which the matrix  $\bl{Q_0}$ is diagonal leads to higher-dimensional generalizations of Kasner spacetimes  \cite{KasnerHD1,KasnerHD2,KasnerHD3,KasnerHD4,KasnerHD4}, which are generally used to model homogeneous but anisotropic cosmological systems \cite{Belinsky:1970ew,Jacobs}.
However, the most interesting cases are the ones in which the matrix $\bl{Q_0}$ cannot be diagonalized, namely when $\bl{Q_0}$ is complex.  As far as the authors know, the latter higher-dimensional solutions have not been described in the literature yet.

The case $n=2$ of the such problem have already been considered recently in Ref. \cite{GabrielBatista2}, as a part of a broader program of integrating Einstein's equation for four-dimensional spacetimes endowed with two commuting Killing vectors and a non-trivial Killing tensor of rank two, see also Refs. \cite{AnabalonBatista,GabrielBatista,BenentiFrancaviglia}. It has been proven there that almost all solutions that arise from integrating Einstein's vacuum equation for the line element (\ref{LineElement1}) when $n=2$ are already known, a particular example being the Kasner metric with cosmological constant \cite{Stephani, EKasner}. Nevertheless, it turns out that one solution for the latter problem had not been described in the literature before its appearance in Ref. \cite{GabrielBatista2}. Differently from Kasner spacetime, in this new solution one of the Killing vectors $\partial_{\sigma_i}$ is not orthogonal to a family of hypersurfaces, so that the line element is non-diagonal when we use the cyclic coordinates $\sigma_i$. Thus, besides leading to a whole class of new higher-dimensional solutions of Einstein's vacuum equation, the problem considered in the present article serves also to shed light over the origin of the new four-dimensional solution obtained in Ref. \cite{GabrielBatista2}.

Since the settlement of General Relativity, it has been increasing the interest on spacetimes with dimension greater than four, specially in the last two decades.  The reasons for studying these spaces are abundant. For instance, the gravity/gauge duality provides a map between field theories in $d$ dimensions and gravitational theories in $d+1$ dimensions, linking the weak coupling regime of one side to the strong coupling regime of the other side \cite{Maldacena:1997re,Horowitz:2006ct,Hubeny:2014bla}. Such tool has been used, for example, to obtain results on strongly coupled quantum chromodynamics, which is a field theory in four dimensions, by performing calculations with a weak gravitational field in five dimensions \cite{Erlich:2005qh}. A particularly exciting illustration are the experimentally verified results on quark-gluon plasma \cite{Finazzo:2014cna,CasalderreySolana:2011us}. Another important motivation for studying higher-dimensional spacetimes is string theory, which, among other things, provides a description of quantum gravity. In order to be consistent, string theory requires spacetime to have $10$ dimensions \cite{Mukhi:2011zz}. Besides these two examples, there are several other theories that seek to explain our Universe through the use of spaces with dimension greater than four, for reviews see \cite{Emparan, Csaki:2004ay}. With these motivations in mind, the solutions presented here shall be of particular application to cosmological models, inasmuch as some of them have spatial homogeneity.

The outline of the article is the following. In the next section we establish the basic notation used throughout the article and introduce a frame in order to compute the components of the Ricci tensor through Cartan's structure equations. Then, in Sec. \ref{Sec.Integration} we fully integrate Einstein's vacuum equation with a cosmological constant, namely $R_{ab}=\Lambda g_{ab}$. The solutions obtained depend on the exponential of a symmetric matrix, which is generally hard to compute explicitly. Thus, in Sec. \ref{Sec.CanonicalForm} we use some theorems of matrix theory along suitable coordinate transformations in order to provide an explicit form for the solutions obtained in the preceding section. Particularly, we stress that the solutions can have different algebraic structures depending on the canonical form of the symmetric matrix.  Finally, in Sec. \ref{Sec.Examples} we work out some examples for the solutions found in Sec. \ref{Sec.Integration}.  Also, the conclusions and perspectives of the article are presented in Sec. \ref{Sec.Conc}.

%%%%%%%%%%%%%%%%%%%%%%%%%%%%%%%%%%%%%%%%%%%%%%%%%%%%%%%%%%%%%%%%%%%%%%%%%%%%%%%%%%%%%%%%%%%%%%%%%%%%%%%%%%%%%%%%%%%
%%%%%%%%%%%%%%%%%%%%%%%%%%%%%%%%%%%%%%%%%%%%%%%%%%%%%%%%%%%%%%%%%%%%%%%%%%%%%%%%%%%%%%%%%%%%%%%%%%%%%%%%%%%%%%%%%%%
%%%%%%%%%%%%%%%%%%%%%%%%%%%%%%%%%%%%%%%%%%%%%%%%%%%%%%%%%%%%%%%%%%%%%%%%%%%%%%%%%%%%%%%%%%%%%%%%%%%%%%%%%%%%%%%%%%%
%%%%%%%%%%%%%%%%%%%%%%%%%%%%%%%%%%%%%%%%%%%%%%%%%%%%%%%%%%%%%%%%%%%%%%%%%%%%%%%%%%%%%%%%%%%%%%%%%%%%%%%%%%%%%%%%%%%
%%%%%%%%%%%%%%%%%%%%%%%%%%%%%%%%%%%%%%%%%%%%%%%%%%%%%%%%%%%%%%%%%%%%%%%%%%%%%%%%%%%%%%%%%%%%%%%%%%%%%%%%%%%%%%%%%%%
%%%%%%%%%%%%%%%%%%%%%%%%%%%%%%%%%%%%%%%%%%%%%%%%%%%%%%%%%%%%%%%%%%%%%%%%%%%%%%%%%%%%%%%%%%%%%%%%%%%%%%%%%%%%%%%%%%%

\section{Introducing a Vielbein and Computing the Curvature}

In what follows, we shall use the frame formalism in order to compute the curvature and integrate Einstein's equations. In this preliminary section we shall define the vielbein and then compute the spin coefficients and use them to calculate the Ricci tensor. Before proceeding, let us establish our index conventions. Indices from the beginning of the alphabet, like $a$, $b$ and $c$, run over all spacetime dimensions, from $1$ to $n+2$, whereas indices $i$, $j$ and $k$,  from the middle of the alphabet, range from $1$ to $n$.

First let us define the functions  $F_x^{ij}$ to be such that $H_x^{ij} = \sum_{k=1}^n F_x^{ki}F_x^{kj}$. Defining the $n\times n$ matrices $\bl{H_x}$ and $\bl{F_x}$ to be the ones whose components are $H_x^{ij}$ and $F_x^{ij}$ respectively, the latter relation means that
$$\bl{H_x} = \bl{F_x}^t \, \bl{F_x} \,, $$
with $\bl{F_x}^t$ standing for the transpose of the matrix $\bl{F_x}$. Since $\bl{H_x}$ is a symmetric matrix, it turns out that such decomposition is always possible, as a consequence of Takagi's factorization \cite{MatrixAnalysis}. In particular, since the matrix $\bl{H_x}$ is invertible (otherwise the line element would be singular), it follows that $\bl{F_x}$ is also invertible. Thus, it is possible to define the matrix $\bl{A_x}$  as
\begin{equation}\label{Ax}
  \bl{A_x} = \bl{F_x}'\,\bl{F_x}^{-1}\,,
\end{equation}
where in the above definition we have used the convention adopted henceforth that a prime over a function means a derivative with respect to its variable. Thus, $\bl{F_x}'$ stands for $\frac{d}{dx}\bl{F_x}$.

Then, defining the frame of 1-forms
\begin{align*}
  e^i &= \sqrt{S_x+S_y} \,F_x^{ij}d\sigma_j\,,  \\
  e^{\hat{x}} &= e^{n+1}  = \sqrt{S_x+S_y} \,\frac{dx}{\Delta_x} \,, \\
  e^{\hat{y}} &=  e^{n+2}  = \sqrt{S_x+S_y} \,\frac{dy}{\Delta_y}\,,
\end{align*}
it follows that the line element is given by
\begin{equation*}\label{LineElement2}
  ds^2 = \sum_{a=1}^{n+2} \, (e^a)^2\,.
\end{equation*}
So, the vielbein $\{e^a\}$ is orthonormal, namely if $\{e_a\}$ is the dual frame of vector fields and $\bl{g}$ is the metric tensor, then $\bl{g}(e_a,e_b) = \delta_{ab}$. Depending on the signature of the metric, the frame might be complex in order to enable the existence of such a basis. Indeed, generally the reality conditions of a frame with specified inner products are intimately connected to the signature of the metric \cite{Trautman}.

Now, since we shall assume the Levi-Civita connection, which is torsion-less, the first Cartan equation reads
\begin{equation}\label{1Cartaneq}
  de^a + \omega^a_{\ph{a}b}\wedge e^b = 0 \,,
\end{equation}
where $\omega^a_{\ph{a}b}$ are 1-forms known as spin coefficients or connection 1-forms. Since our connection is compatible with the metric and the components of the metric on the frame $\{e^a\}$ are constant, it follows that $\omega^a{}_b=-\omega^b{}_a$. Equation \eqref{1Cartaneq} can be solved for the spin coefficients $\omega^a{}_b$, to be given in terms of the functions $S_x$, $S_y$, $\Delta_x$, $\Delta_y$ and $F^{ij}_x$. After some algebra, we eventually arrive at the following solution:
\begin{align*}
\omega^i{}_j&= - \frac{\Delta_x(A^{ij}_x-A^{ji}_x)}{2\sqrt{S_x+S_y}} \, e^{\hat{x}},  \\
\omega^i{}_{\hat{x}}&= \Delta_x\Big[\frac{(A^{ij}_x+A^{ji}_x)}{2\sqrt{S_x+S_y}}+\frac{S'_x\delta^i_j}{2(S_x+S_y)^{3/2}}\Big] \,e^j, \\
\omega^i{}_{\hat{y}}&=\frac{\Delta_y S'_y}{2(S_x+S_y)^{3/2}}  \, e^i, \\
\omega^{\hat{x}}{}_{\hat{y}}&=\frac{(\Delta_y S'_y \,e^{\hat{x}}-\Delta_x S'_x \,e^{\hat{y}} )}{2(S_x+S_y)^{3/2}}\,,
\end{align*}
where $A_x^{ij}$ stands for the components of the matrix $\bl{A_x}$ defined in Eq. (\ref{Ax}).

With these spin coefficients at hand, one can straightforwardly compute the components of the curvature tensor by means of the second Cartan equation, which is given by
\begin{equation*}
\frac{1}{2}\,R^a_{\ph{a}bcd}\,e^c\wedge e^d = d\omega^a{}_b+\omega^a{}_c\wedge\omega^c{}_b\,,
\end{equation*}
where $R^a_{\ph{a}bcd}$ stands for the components of the Riemann tensor in the frame. Performing these calculations and then computing the components of the Ricci tensor $R_{ab} = R^c_{\ph{c}acb}$, we eventually arrive at the following expressions
\begin{widetext}
\begin{align}
R_{ij}&=\frac{-\Delta_x^2}{2(S_x+S_y)} \left[ \left(tr(\bl{A_x}) + \frac{n S_x'}{2(S_x+S_y)} + \frac{\Delta_x'}{\Delta_x}\right) ( \bl{A_x} +\bl{A_x}^t)-\bl{A_x}\bl{A_x}^t+\bl{A_x}^t \bl{A_x} + \bl{A_x}' +\bl{A_x}'^{\,t} \,\right]_{ij} \nonumber\\
 &-\frac{\delta_{ij}}{2(S_x+S_y)^2}\left[ \Delta_x^2 \left( tr(\bl{A_x}) \, S_x' + S_x'' + \frac{(n-2) (S'_x)^2}{2(S_x+S_y)} +  \frac{S_x'\Delta'_x }{\Delta_x } \right) +
 \Delta_y^2 \left(  S_y'' + \frac{(n-2) (S'_y)^2}{2(S_x+S_y)} +   \frac{S_y'\Delta'_y }{\Delta_y }  \right)  \right]\,, \nonumber\\
 \nonumber\\
R_{\hat{x}\hat{x}}&=\frac{-\Delta_x^2}{2(S_x+S_y)} \left[\, tr(\bl{A_x} \bl{A_x}^t) + tr(\bl{A_x}^2) + \left(\frac{S'_x}{S_x+S_y} + \frac{2\Delta'_x}{\Delta_x} \right) tr(\bl{A_x})+ 2tr(\bl{A_x}') \, \right] \nonumber\\
&-\frac{(n+1)\Delta_x^2}{2(S_x+S_y)^2}\left[ S''_x - \frac{(S'_x)^2}{S_x+S_y} + \frac{S'_x\Delta'_x}{\Delta_x} \right] - \frac{\Delta_y^2}{2(S_x+S_y)^2} \left[ S''_y + \frac{(n-2)(S'_y)^2}{2(S_x+S_y)} + \frac{S'_y\Delta'_y}{\Delta_y} \right]\,,  \label{Ricci}\\
 \nonumber\\
R_{\hat{y}\hat{y}}&=\frac{-\Delta_x^2}{2(S_x+S_y)^2} \left[tr(\bl{A_x})\,S'_x + S''_x + \frac{(n-2)(S'_x)^2}{2(S_x+S_y)}  + \frac{S'_x\Delta'_x}{\Delta_x}\, \right]
+\frac{(n+1)\Delta_y^2}{2(S_x+S_y)^2} \left[\,\frac{(S'_y)^2}{S_x+S_y} - S''_y  - \frac{S'_y\Delta'_y}{\Delta_y}\, \right] \,, \nonumber \\
 \nonumber\\
R_{\hat{x}\hat{y}}&=\,\frac{3n \Delta_x \Delta_y S'_xS'_y}{4(S_x+S_y)^3}\,, \qquad R_{i\hat{x}}=0\,, \qquad R_{i\hat{y}}=0. \nonumber
\end{align}
\end{widetext}
In these expressions, $tr(\bl{A_x})$ denotes the trace of the matrix $\bl{A_x}$.
At this point, it is worth mentioning that, since our frame is orthonormal, the frame indices can be risen or lowered without changing the value of the component. Thus, $R^{a}_{\ph{a}b} = R_{ab} = R^{b}_{\ph{b}a}$, where the symmetry of the Ricci tensor has been used in the last equality.

%%%%%%%%%%%%%%%%%%%%%%%%%%%%%%%%%%%%%%%%%%%%%%%%%%%%%%%%%%%%%%%%%%%%%%%%%%%%%%%%%%%%%%%%%%%%%%%%%%%%%%%%%%%%%%%%%%%
%%%%%%%%%%%%%%%%%%%%%%%%%%%%%%%%%%%%%%%%%%%%%%%%%%%%%%%%%%%%%%%%%%%%%%%%%%%%%%%%%%%%%%%%%%%%%%%%%%%%%%%%%%%%%%%%%%%
%%%%%%%%%%%%%%%%%%%%%%%%%%%%%%%%%%%%%%%%%%%%%%%%%%%%%%%%%%%%%%%%%%%%%%%%%%%%%%%%%%%%%%%%%%%%%%%%%%%%%%%%%%%%%%%%%%%
%%%%%%%%%%%%%%%%%%%%%%%%%%%%%%%%%%%%%%%%%%%%%%%%%%%%%%%%%%%%%%%%%%%%%%%%%%%%%%%%%%%%%%%%%%%%%%%%%%%%%%%%%%%%%%%%%%%
%%%%%%%%%%%%%%%%%%%%%%%%%%%%%%%%%%%%%%%%%%%%%%%%%%%%%%%%%%%%%%%%%%%%%%%%%%%%%%%%%%%%%%%%%%%%%%%%%%%%%%%%%%%%%%%%%%%
%%%%%%%%%%%%%%%%%%%%%%%%%%%%%%%%%%%%%%%%%%%%%%%%%%%%%%%%%%%%%%%%%%%%%%%%%%%%%%%%%%%%%%%%%%%%%%%%%%%%%%%%%%%%%%%%%%%

\section{Integration of Einstein's Vacuum Equation}\label{Sec.Integration}

In this section we will fully integrate Einstein's field equation in vacuum with a cosmological constant $\Lambda$. Namely, we shall attain the most general solution for the equation
\begin{equation*}
R_{ab}=\Lambda\, \delta_{ab}\,.
\end{equation*}
In particular, the component $\hat{x}\hat{y}$ of the latter equation implies $R_{\hat{x}\hat{y}}=0$. Thus, looking at Eq. (\ref{Ricci}), we conclude that the combination $\Delta_x \Delta_y S'_xS'_y$ should be identically zero. Since $\Delta_x$ and $\Delta_y$ cannot be zero (otherwise the line element (\ref{LineElement1}) would be meaningless), it follows that $S_x'S_y'$ must vanish. Therefore, we have two possibilities:  (A) $S_y$ is constant;  (B) $S_x$ is constant and $S_y$ non-constant. In particular, note that case (A) encompasses the possibility of both functions $S_x$ and $S_y$ being constant as well as the case in which just $S_y$ is constant, with $S'_x$ not identically zero. In what follows, we shall tackle the cases (A) and (B) separately.
\vspace{0.5cm}

\subsection{The Case $S'_y=0$}

In this subsection we will integrate Einstein's equation for the case  (A), namely when $S_y$ is constant. Since $S_y$ shows up in the metric only through the combination $S_x+S_y$,  we can absorb the constant value of $S_y$ into $S_x$ and assume that $S_y$ vanishes.  We shall also redefine the coordinate $y$ in such a way to eliminate the dependence on the function $\Delta_y$ of this line element. More precisely, we will replace the coordinate $y$ by $\tilde{y} = \int dy/\Delta_y$. Dropping the tilde after the transformation, this amounts to assuming $\Delta_y=1$ in the line element (\ref{LineElement1}).
Therefore, in this subsection we shall set
\begin{equation}\label{SyDy}
  S_y = 0 \;\;\textrm{ and }\;\; \Delta_y = 1 \,,
\end{equation}
which represent no loss of generality. Inserting these choices into (\ref{Ricci}), it follows that the non-vanishing components of the Ricci tensor are
\begin{widetext}
\begin{align}
R_{\hat{x}\hat{x}}&=\frac{-\Delta_x^2}{2S_x} \left[\, tr(\bl{A_xA_x}^t) + tr(\bl{A_x}^2) + \left(\frac{S'_x}{S_x} + \frac{2\Delta'_x}{\Delta_x} \right) tr(\bl{A_x})+ 2tr(\bl{A_x}')
+\frac{(n+1)}{S_x}\left( S''_x - \frac{(S'_x)^2}{S_x} + \frac{S'_x\Delta'_x}{\Delta_x} \right) \right] \,, \nonumber \\
\nonumber\\
R_{\hat{y}\hat{y}}&=\frac{-\Delta_x^2}{2S_x^2} \left[tr(\bl{A_x})\,S'_x + S''_x + \frac{(n-2)(S'_x)^2}{2S_x}  + \frac{S'_x\Delta'_x}{\Delta_x} \right] \,, \label{RicciA}\\
\nonumber\\
R_{ij}&=\frac{-\Delta_x^2}{2S_x} \left[ \left(tr(\bl{A_x}) + \frac{n S_x'}{2S_x} + \frac{\Delta_x'}{\Delta_x}\right)(\bl{A_x}+\bl{A_x}^t)-\bl{A_xA_x}^t+\bl{A_x}^t \bl{A_x} + \bl{A_x}' +\bl{A_x}'^{\,t} \,\right]_{ij}+\delta_{ij}R_{\hat{y}\hat{y}}\,,\nonumber
\end{align}
\end{widetext}

Now, let us impose Einstein's equation $R_{ab}=\Lambda \delta_{ab}$.  Due to the assumption (\ref{SyDy}), which guarantees $R_{\hat{x}\hat{y}}=0$, and since $R_{i\hat{x}}$ and $R_{i\hat{y}}$ are already zero for the class of spaces considered here, the only parts of Einstein's equation that need to be demanded
are $R_{ij}=\Lambda\delta_{ij}$,   $R_{\hat{x}\hat{x}}=\Lambda $, and $R_{\hat{y}\hat{y}}=\Lambda $.
Hence, the left hand side of the last equation in (\ref{RicciA}) can be equated to $\Lambda \delta_{ij}$, which cancels the term $R_{\hat{y}\hat{y}}\delta_{ij}$ on the right hand side of this equation, thus yielding
\begin{multline}\label{MatrixRel1}
 \left[tr(\bl{A_x}) + \frac{n S_x'}{2S_x} + \frac{\Delta_x'}{\Delta_x}\right](\bl{A_x}+\bl{A_x}^t)\\
-\bl{A_xA_x}^t+\bl{A_x}^t\bl{A_x} + \bl{A_x}'+\bl{A_x}'^{\,t}=0\,.
\end{multline}
Therefore, the problem of solving Einstein's equation for the case $S'_y=0$ reduces to setting $R_{\hat{x}\hat{x}}=\Lambda$ and $R_{\hat{y}\hat{y}}=\Lambda$ in Eq. \eqref{RicciA}, along solving with the matrix equation \eqref{MatrixRel1}.

A further simplification can be accomplished by noting that the function $\Delta_x$ can be chosen as desired by means of redefining the coordinate $x$.   Therefore, without loss of generality, let us choose $\Delta_x$ to be the function that makes the expression enclosed by the square brackets in Eq. \eqref{MatrixRel1} identically zero, namely let us assume
\begin{equation}\label{DeltaConstraint}
tr(\bl{A_x}) + \frac{n S_x'}{2S_x} + \frac{\Delta_x'}{\Delta_x}=0\,.
\end{equation}
With this gauge choice, Eq. (\ref{MatrixRel1}) boils down to
\begin{equation}\label{MatrixRel2}
\bl{A_x}' +\bl{A_x}'^{\,t}-\bl{A_xA_x}^t+\bl{A_x}^t \bl{A_x} =0\,.
\end{equation}
Then, by taking the trace of this equation, we find out that, in these coordinates, the trace of the matrix $\bl{A_x}$ is constant. We shall denote this constant by $a_1$
\begin{equation}\label{tr1}
  tr(\bl{A_x}) = a_1\,.
\end{equation}
With this in mind, we are now able to solve equation \eqref{DeltaConstraint}, the general solution being given by
\begin{equation}\label{Deltax1}
\Delta_x=c_1 e^{-a_1 x}(S_x)^{-n/2}\,,
\end{equation}
where $c_1$ is an integration constant. Likewise, multiplying Eq. \eqref{MatrixRel2} on the left by $\bl{A_x}$ and taking the trace of the resulting equation, we find that $tr(\bl{A_x}^2+\bl{A_x A_x}^t)$ is a constant, conveniently denoted here by $2a_2/(n+1)$. Thus, we have
\begin{equation}\label{tr2}
tr(\bl{A_x}^2+\bl{A_x A_x}^t)=\frac{2}{n+1}\,a_2\,.
\end{equation}
This piece of information is particularly useful since the latter combination appears in the expression for $R_{\hat{x}\hat{x}}$ in Eq. \eqref{RicciA}.
Using Eqs. \eqref{tr2} and \eqref{RicciA}, it follows from the integration of $R_{\hat{x}\hat{x}}-R_{\hat{y}\hat{y}}=0$ that
\begin{equation}\label{Sx}
  S_x = s_1\!\left[e^{a_1x}\cosh\!\left(\! \sqrt{\frac{a_2-a_1^2}{n}}(x-x_0)\!\right)\right]^{\frac{-2}{n+1}}\!\!\!\!,
\end{equation}
where $s_1$ and $x_0$ are new integration constants.
Then, inserting Eqs. \eqref{Deltax1} and \eqref{Sx} into the equation $R_{\hat{y}\hat{y}}=\Lambda$, we find that the constant $a_2$ must be related to $s_1$, $\Lambda$, $c_1$ and $a_1$ through the following relation:
\begin{equation}\label{c2}
  a_2= \frac{1}{c_1^2}\left[c_1^2a_1^2+n(n+1)s_1^{n+1}\Lambda \right]\,.
\end{equation}

%\textbf{Carlos Parou AQUI}

Now, it only remains to solve the matrix equation \eqref{MatrixRel2}. In order to accomplish this task, let us first define the  matrices $\bl{Q_x}=\frac{1}{2}(\bl{A_x}+\bl{A_x}^t)$ and $\bl{P_x}=\frac{1}{2}(\bl{A_x}-\bl{A_x}^t)$, which represent independent degrees of freedom of the matrix $\bl{A_x}$, namely its symmetric and anti-symmetric parts respectively. By means of such definitions, equation \eqref{MatrixRel2} can be written as
\begin{equation}\label{LaxPair}
\bl{Q_x}'=[\bl{P_x},\bl{Q_x}]\,,
\end{equation}
where the symbol $[\,\,,\,]$ stands for the ordinary matrix commutator. This equation is a Lax pair equation, which is generally connected to integrable systems and leads to an infinity amount of conserved charges \cite{Lax68,LaxReview,CarigliaLax}. In particular, since the trace of a commutator is zero, it follows from Eq. (\ref{LaxPair}) that $tr(\bl{Q_x})$ is a constant. More generally, the trace of an arbitrary power of $\bl{Q_x}$ must be constant, which stems from the relation
\begin{equation*}\label{LaxPair_p}
\frac{d}{dx}(\bl{Q_x}^p)=[\bl{P_x},\bl{Q_x}^p]\,.
\end{equation*}
%from which one concludes that $tr(\bl{Q_x}^p)=constant$.

%Rather than finding $\bl{A_x}$ and $\bl{F_x}$, we are mainly interested in obtaining  $\bl{H_x}= \bl{F_x}^t \bl{F_x}$ , which is the matrix appearing on the line element \eqref{LineElement1}. Bearing this in mind, let us find a differential equation for $\bl{H_x}$. From the very definition of $\bl{A_x}$, see Eq. (\ref{Ax}), it follows that

In order to continue the integration process, notice that rather than finding $\bl{A_x}$ and $\bl{F_x}$, we are mainly interested in obtaining  $\bl{H_x}= \bl{F_x}^t \bl{F_x}$, which is the matrix appearing on the line element \eqref{LineElement1}. Bearing this in mind, let us find a differential equation for $\bl{H_x}$. From the very definition of $\bl{A_x}$, see Eq. (\ref{Ax}), it follows that
\begin{equation*}
  \frac{d}{dx} \bl{F_x} = \bl{A_x F_x} =(\bl{Q_x}+\bl{P_x}) \bl{F_x} \,.
\end{equation*}
Taking the transpose of the latter relation, it follows that
\begin{equation*}
\frac{d}{dx} \bl{F_x}^t = \bl{F_x}^t(\bl{Q_x}-\bl{P_x}) \,.
\end{equation*}
Thus, differentiating  $\bl{H_x}=\bl{F_x}^t \bl{F_x}$  and using these relations along with (\ref{LaxPair}), we eventually arrive at
\begin{equation*}
  \frac{d^p}{dx^p} \bl{H_x} = 2^p \bl{F_x}^t \bl{Q_x}^p \bl{F_x} \,.
\end{equation*}
Therefore, all the derivatives of $\bl{H_x}$ at one point are determined by the values of $\bl{F_x}$ and $\bl{Q_x}$ at this single point. In particular,
denoting the values of  $\bl{F_x}$ and $\bl{Q_x}$ at $x=0$ by $\bl{F_0}$ and $\bl{Q_0}$ respectively, it follows that
\begin{equation*}
  \left. \frac{d^p}{dx^p} \bl{H_x}\,\right|_{x=0} = 2^p \bl{F_0}^t \bl{Q_0}^p \bl{F_0} \,,
\end{equation*}
so that the Taylor series of $\bl{H_x}$ reads
\begin{equation}\label{Hx1}
  \bl{H_x}(x) = \bl{F_0}^t\,\sum_{p=0}^{\infty} \frac{(2x)^p}{p!}\,\bl{Q_0}^p \, \bl{F_0} = \bl{F_0}^t\,e^{2x\bl{Q_0}}\,\bl{F_0} \,.
\end{equation}
From this, we have that the line element \eqref{LineElement1}, for the case $S'_y=0$, can be written as
\begin{equation*}
  ds^2 = S_x\left[ \bl{d\sigma}^t \,\bl{F_0}^t\,e^{2x \bl{Q_0}}\,\bl{F_0} \,\bl{d\sigma} + \frac{dx^2}{\Delta_x^2} + dy^2  \right] \,,
\end{equation*}
where $\bl{d\sigma}$ is the $n\times 1$ matrix whose components are the $d\sigma_i$. Actually, since $\bl{F_x}$ is invertible, it follows that the dependence on $\bl{F_0}$ can be gauged away by defining the cyclic coordinates $\tilde{\sigma}_i$ as
\begin{equation}\label{CoordTranSigma}
\tilde{\sigma}_i = F_0^{ij}\,\sigma_j\,.
\end{equation}
In terms of these, the final line element is given by
\begin{equation}\label{LineElement3}
ds^2 = S_x\left[ \bl{d\tilde{\sigma}}^t \,e^{2x \bl{Q_0}}\,\bl{d\tilde{\sigma}} + \frac{dx^2}{\Delta_x^2} + dy^2  \right] \,.
\end{equation}
In this final solution, the functions $\Delta_x$ and $S_x$ are given respectively by \eqref{Deltax1} and \eqref{Sx}, whereas $\bl{Q_0}$ is an arbitrary constant symmetric matrix. Since the integration constants $a_1$ and $a_2$ appearing in the expressions for $\Delta_x$ and $S_x$  are related to the traces of $\bl{A_x}$ and $(\bl{A_x}^2 + \bl{A_x} \bl{A_x}^t)$, see Eqs. (\ref{tr1}) and (\ref{tr2}), it follows that they must be determined by the traces of $\bl{Q_0}$ and its powers. Indeed, one can check that they must be given by
\begin{equation}\label{a0c2}
a_1 = tr(\bl{Q_0}) \quad \text{and} \quad a_2 = (n+1)\,tr(\bl{Q_0}^2) \,.
\end{equation}
This is the most general solution for Einstein's equation in the case $S'_y=0$.

In particular, from the expression for $S_x$, \eqref{Sx}, we see that
the case in which $S_x$ and $S_y$ are both constant can be attained by setting $a_1$
and $a_2$ to zero.
Nevertheless, going through the whole integration process assuming $S_x' = 0$ from the very beginning,
one can check that this last requirement is, actually,
not necessary. Rather, the constants $a_1$ and  $a_2$ are just constrained by
the relation $a_2 = (n+1)a_1^2$. Thus, the case  $a_2 = a_1 = 0$ is just a particular
solution. So, the most general solution for this case is provided by the metric
\eqref{LineElement3} with $S_x=s_1$, $\Delta_x$ given by \eqref{Deltax1},
and $\bl{Q_0}$ satisfying $tr(\bl{Q_0})^2=tr(\bl{Q_0}^2)$. Moreover, one can easily see from the condition
$R_{\hat{y}\hat{y}}=\Lambda$ that if $S_x$ and $S_y$ are both constant one must have $\Lambda=0$ in order to attain a solution, see \eqref{RicciA}.

\subsection{Subcase $S'_x=0$ and $S'_y\neq0$}\label{SubsectionSx}

Now, let us consider the subcase in which $S_x$ is constant while $S_y$ is a nonconstant function of $y$. Here we follow steps  analogous to the ones taken in the previous subsection. For instance, absorbing the constant value of $S_x$ into $S_y$, we can assume, without any loss of generality, that
$$S_x=0  \,. $$
With this choice, the nonzero components of the Ricci tensor are
\begin{widetext}
\begin{align*}
R_{\hat{x}\hat{x}}&=\frac{-\Delta_x^2}{2S_y} \left[\, tr(\bl{A_x}\bl{A_x}^t) + tr(\bl{A_x}^2) + \left(\frac{2\Delta'_x}{\Delta_x} \right) tr(\bl{A_x})+ 2tr(\bl{A_x}') \, \right]
- \frac{\Delta_y^2}{2S_y^2} \left[ S''_y + \frac{(n-2)(S'_y)^2}{2S_y} + \frac{S'_y\Delta'_y}{\Delta_y} \right]\,, \\
\\
R_{\hat{y}\hat{y}}&=\frac{-(n+1)\Delta_y^2}{2S_y^2} \left[\,S''_y-\frac{(S'_y)^2}{S_y} + \frac{S'_y\Delta'_y}{\Delta_y}\, \right] \,,\\
\\
R_{ij}&=\frac{-\Delta_x^2}{2S_y} \left[ \left(tr(\bl{A_x})+ \frac{\Delta_x'}{\Delta_x}\right)(\bl{A_x}+\bl{A_x}^t)-[\bl{A_x},\bl{A_x}^t]+ \bl{A_x}' +\bl{A_x}'^t \,\right]_{ij}
\!\!\!\!-\frac{\Delta_y^2\delta_{ij}}{2S_y^2}\left[ S_y'' + \frac{(n-2) (S'_y)^2}{2S_y} +   \frac{S_y'\Delta'_y }{\Delta_y } \right]\,.
\end{align*}
\end{widetext}

Now, recall that the functions $\Delta_x$ and $\Delta_y$ can be chosen arbitrarily, which is equivalent to performing coordinate transformations in $x$ and $y$ respectively. Thus, let us conveniently choose them to be such that the following equations hold
\begin{align}
&\frac{\Delta_x'}{\Delta_x}+tr(\bl{A_x})=0\,,\label{ChoiceDeltax} \\
&\frac{\Delta_y^2}{2S_y^2}\left[ S_y'' + \frac{(n-2) (S'_y)^2}{2S_y} +   \frac{S_y'\Delta'_y }{\Delta_y } \right]=-\Lambda\,.\label{ChoiceDeltay}
\end{align}
With these choices, it follows that the part $R_{ij}=\Lambda \delta_{ij}$ of Einstein's equation reduces to
\begin{equation*}\label{MatrixRel3}
\frac{d}{dx}(\bl{A_x} + \bl{A_x}^{t}) =  [ \bl{A_x},  \bl{A_x}^t] \,,
\end{equation*}
which is the same matrix equation that we have addressed in the previous subsection, with the general solution for $\bl{H_x}$ being given by $\eqref{Hx1}$.
In particular, the traces of  $\bl{A_x}$ and $(\bl{A_x}^2+\bl{A_x}\bl{A_x}^t)$ are constant. Thus, let us define the constants $a_1$ and $a_2$  as follows
\begin{equation}\label{a1a2Sx}
  tr(\bl{A_x}) = a_1 \;\; \textrm{and} \;\; tr(\bl{A_x}^2+\bl{A_x A_x}^t)=\frac{2}{n+1}a_2 \,.
\end{equation}
Taking this into account, it follows that the general solutions for (\ref{ChoiceDeltax}) and (\ref{ChoiceDeltay}) are
\begin{equation}\label{DxDyB}
  \left\{
    \begin{array}{ll}
      \Delta_x(x)=c_1 e^{-a_1x}\,, \\
      \quad  \\
      \Delta_y=-\frac{S_y }{S_y'} \sqrt{\frac{d_1}{(S_y)^n}-\frac{4\Lambda S_y}{n+1}}  \,,
    \end{array}
  \right.
\end{equation}
where $c_1$ and $d_1$ are arbitrary integration constants.

Now, the equations $R_{\hat{x}\hat{x}}=\Lambda$ and $R_{\hat{y}\hat{y}}=\Lambda$ imply the following constraints on the integration constants:
\begin{equation}\label{c2c1}
  a_2=(n+1)a_1^2  \quad \textrm{and} \quad  d_1=0 \,.
\end{equation}
Then, performing the coordinate transformation (\ref{CoordTranSigma}) on the cyclic coordinates, it follows that the general solution of Einstein's equation for the case considered in the present subsection is given by
\begin{align}
 ds^2 = S_y & \Big[ \bl{d\tilde{\sigma}^t} \,e^{2x \bl{Q_0}}\,\bl{d\tilde{\sigma}} + e^{2a_1 x} dx^2 \nonumber \\
 &\quad - \frac{(n+1)(S_y')^2}{4 \Lambda (S_y)^3} dy^2 \Big] \,, \label{LineElement4}
\end{align}
where we have set $c_1 = 1$, a choice that can be accomplished by means of a translation on the coordinate $x$ along with a coordinate transformation in the cyclic coordinates. Note that no restriction has been placed over the function $S_y$, conveying the fact that we still have a coordinate freedom to choose $S_y$ to be any non-constant function of $y$, each choice corresponding to a different coordinate system but representing the same physical spacetime. In the latter line element, the matrix $\bl{Q_0}$ is an arbitrary $n\times n$ symmetric matrix such that
\begin{equation*}
 tr(\bl{Q_0})=a_1 \quad \text{and} \quad tr(\bl{Q_0}^2)=a_1^2\,,
\end{equation*}
where the last constraint stems from Eqs. (\ref{a1a2Sx}) and (\ref{c2c1}). This is the general solution for Einstein's vacuum equation when $S_x'=0$, with $S_y'$ non-vanishing.

\section{Putting the Solution in a Treatable form}\label{Sec.CanonicalForm}

Although the solutions \eqref{LineElement3} and \eqref{LineElement4} have been expressed in an elegant and compact form, the exponential $e^{2x\bl{Q_0}}$ will mostly result in an extremely cumbersome matrix, specially for solutions in dimension higher than four, namely when the values of $n$ are greater than two. Actually, no general expression exists for the exponential of an $n\times n$ matrix when $n>2$. In spite of this, the intent of the present section is to show that, by means of a suitable choice of cyclic coordinates, the term $\bl{d\tilde{\sigma}}^t e^{2x\bl{Q_0}} \bl{d\tilde{\sigma}}$ can always be explicitly written as a finite sum of terms.

%Thus, the present section is devoted to working on the exponential part $\bl{d\tilde{\sigma}^t} e^{2x\bl{Q_0}} \bl{d\tilde{\sigma}}$ of the solutions found %above, in order to make them treatable to practical purposes.

Recall that the matrix $\bl{Q_0}$ appearing on the exponential must be symmetric. Therefore, due to the spectral theorem, it follows that if $\bl{Q_0}$ is real then it can be put in a diagonal form by means of an orthogonal transformation on the basis. Once put in the diagonal form, it is trivial to compute its exponential. However, the case in which $\bl{Q_0}$ is complex is trickier. Fortunately, there exists a complementary theorem for complex symmetric matrices. It states that if the complex symmetric matrix $\bl{Q_0}$ can be diagonalized this will be accomplished by a complex orthogonal transformation \cite{MatrixAnalysis}. Moreover, in the case in which $\bl{Q_0}$ cannot be diagonalized, there always exist an orthogonal complex matrix $\bl{M}$ such that $\bl{M}^{-1} \bl{Q_0 M}$ is the sum of a nilpotent matrix plus a diagonal matrix that commutes with it, so that the exponential can also be easily computed \cite{Matrix1,Matrix2}. By ``orthogonal complex matrix'' we mean an $n\times n$ matrix $\bl{M}$ with complex entries obeying the relation $\bl{M}^t \bl{M}=\bl{I_n}$, where $\bl{I_n}$ stands for the $n\times n$ identity matrix. In what follows we shall consider the cases in which $\bl{Q_0}$ is diagonalizable or not separately.

\subsection{Diagonalizable Case}

In this subsection we consider the symmetric matrix $\bl{Q_0}$ to be diagonalizable. Thus, according to the theorem described above, there exits an orthogonal matrix $\bl{M}$ such that
\begin{equation*}
  \bl{M}^{-1} \bl{Q_0 M} = \textrm{diag}(q_1,q_2,\cdots,q_n)\,,
\end{equation*}
where $q_i$ are the eigenvalues of $\bl{Q_0}$ \cite{MatrixAnalysis}.
Following this reasoning, let us perform the change of coordinates $\tilde{\sigma}_i \rightarrow \tau_i$ defined by
\begin{equation}\label{tau}
 \bl{\tilde{\sigma}} =  \bl{M} \cdot  \bl{\tau}\,,
\end{equation}
where $\bl{\tau}$ is an $n\times 1$ matrix with entries $\tau_i$. Additionally, let us use the identity $\bl{M}^{-1} e^{\bl{Q}}\bl{M}=e^{ \bl{M}^{-1}\bl{Q M}}$, which always holds. Then, since $\bl{M}$  is orthogonal, it follows that $\bl{M}^{-1}=\bl{M}^t$, so that we can write
\begin{align*}
 \bl{d\tilde{\sigma}}^t e^{2x\bl{Q_0}} \bl{d\tilde{\sigma}} &= \bl{d\tau}^t \bl{M}^{t} e^{2x\bl{Q_0}}  \bl{M} \bl{d\tau} \\
 &= \bl{d\tau}^t e^{2x \bl{M}^t \bl{Q_0}  \bl{M}} \bl{d\tau} =\sum_{i=1}^{n} e^{2x q_i} (d\tau_{i})^2 \,.
\end{align*}
Therefore, in the case in which $\bl{Q_0}$ is diagonalizable, the line elements  \eqref{LineElement3} and \eqref{LineElement4} also become diagonal,
\begin{equation}\label{LineElement5}
  ds^2 = (S_x + S_y)\left[ \sum_{i=1}^{n} e^{2x q_i} d\tau_{i}^2 + \frac{dx^2}{\Delta_x^2} + \frac{dy^2}{\Delta_y^2}  \right] \,,
\end{equation}
where the functions $S_x$, $S_y$, $\Delta_x$ and $\Delta_y$ should be the ones found in the previous section while integrating Einstein's vacuum equation.

\subsection{Nondiagonalizable Case}\label{Sec.NonDiagTheory}

In this subsection we shall consider the case in which $\bl{Q_0}$ is not diagonalizable. We begin by stating the result that for any complex symmetric matrix $\bl{Q_0}$ there exists a complex orthogonal matrix $\bl{M}$ such that $\bl{M}^{-1}\bl{Q_0 M}$ is block-diagonal, with each block possessing a canonical form \cite{Matrix1,Matrix2}. More explicitly, we have
\begin{equation*}
  \bl{M}^{-1}\bl{Q_0} \bl{M}=\bl{Q_1}\oplus \bl{Q_2}\oplus\cdots\oplus \bl{Q_\ell} \,,
\end{equation*}
where $\bl{Q_\nu}$ is an $m_{\nu}\times m_\nu$ matrix given by
\begin{equation*}
  \bl{Q_\nu} = q_\nu \, \bl{I_{m_\nu}} + \bl{P_{m_\nu,0}} \,.
\end{equation*}
In the latter expression $\bl{I_{m_\nu}}$ stands for the $m_{\nu}\times m_\nu$ identity matrix, $q_\nu$ is an eigenvalue of $\bl{Q_0}$, whereas $\bl{P_{m_\nu,0}}$ is a nilpotent matrix described in the sequel. The index $\nu$ range from 1 to $\ell$, with $\ell$ being the number of blocks in the canonical form of $\bl{Q_0}$.

In order to define $\bl{P_{m,0}}$ it is useful to define the set of $m\times m$ matrices $\bl{P_{m,k}}$, with $k$ being an integer, as the matrices whose components are given by
\begin{align}
P_{m,p}^{ij}=\frac{1}{2}\Big[&(\delta_{i,j+p+1}+\delta_{i,j-p-1})\Big.\nonumber\\
&\Big.+i(\delta_{i,m-p-j}-\delta_{i,m+p-j+2})\Big]\, ,\label{Pnk}
\end{align}
where the indices $i$ and $j$ range from 1 to $m$. As can easily be checked, these matrices have the following properties:
\begin{equation}\label{P}
  \left\{
    \begin{array}{ll}
      (\bl{P_{m,0}})^{p+1}=\bl{P_{m,p}}\,,  \\
\quad \\
     \bl{P_{m,p}}=0  \; \textrm{if} \;  p\geq m-1 \,.
    \end{array}
  \right.
\end{equation}
In particular, note that the matrix $\bl{P_{m,0}}$ is nilpotent, with index $m$.  Moreover, note that $\bl{P_{1,0}}=0$, so that if the block
$\bl{Q_\nu} = q_\nu \bl{I_{m_\nu}} + \bl{P_{m_\nu,0}}$ is a $1\times 1$ matrix it becomes just the eigenvalue $q_\nu$.

With these tools at hand, we are now ready to compute the exponential $e^{2x \bl{Q_0}}$. First, since $\bl{M}$ is an orthogonal matrix we have that $\bl{M}^{-1} = \bl{M}^t$. Therefore, by defining the coordinates $\tau_i$ just as we did in Eq. (\ref{tau}), it follows that
\begin{align*}
  \bl{d\tilde{\sigma}}^t e^{2x\bl{Q_0}} \bl{d\tilde{\sigma}} &=  \bl{d\tau}^t \bl{M}^t\,e^{2x \bl{Q_0}}\,  \bl{M}\bl{d\tau}\\
&=  \bl{d\tau}^t \bl{M}^{-1}\,e^{2x \bl{Q_0}}\,  \bl{M}\bl{d\tau}\\
&=  \bl{d\tau}^t \,e^{2x \bl{M}^{-1}\bl{Q_0}\bl{M}}\,  \bl{d\tau}\\
&=  \bl{d\tau}^t \,\left(e^{2x\bl{Q_1}}\oplus \cdots\oplus e^{2x\bl{Q_\ell}}\right)\,  \bl{d\tau}\\
&=  \sum_{\nu=1}^{\ell}\,\bl{d\tau_\nu}^t \,e^{2x\bl{Q_\nu}}\, \bl{d\tau_\nu} \,,
\end{align*}
where $\bl{d\tau_\nu}$ is a $m_{\nu}\times 1$ column matrix. Thus, the problem of computing the exponential $e^{2x \bl{Q_0}}$ has been reduced to the task of calculating $e^{2x\bl{Q_\nu}}$. At first glance it may seem that we have accomplished nothing, since we still have to exponentiate a matrix, although generally smaller. Nevertheless, now these matrices $\bl{Q_\nu}$ have a canonical form that greatly facilitate the job. Indeed, due to the property (\ref{P}) it follows that $(\bl{P_{m_\nu,0}})^{m_\nu} = 0$, so that in the exponential series $e^{2x\bl{P_{m_\nu,0}}}$  we just need to consider terms up to the power $(\bl{P_{m_\nu,0}})^{m_\nu-1}$. Thus, since $\bl{Q_\nu} = q_\nu \bl{I_{m_\nu}} + \bl{P_{m_\nu,0}}$ and since the identity matrix $\bl{I_{m_\nu}}$ commutes with every other $m_\nu\times m_\nu$ matrix, we can conclude that
\begin{align}
  e^{2x\bl{Q_\nu}} & = e^{2x q_\nu \bl{I_{m_\nu}}} e^{2x\bl{P_{m_\nu,0}}} \nonumber\\
   & =  e^{2x q_\nu } \sum_{p=0}^{m_\nu-1}\frac{(2x)^p}{p!}\bl{P_{m_\nu,p-1}} \,. \label{ExpQ}
\end{align}
Therefore, when $\bl{Q_0}$ cannot be diagonalized the final form of our line element becomes:
\begin{align}\label{Nondiagmetric}
  &ds^2 = (S_x + S_y)\bigg[ \frac{dx^2}{\Delta_x^2} + \frac{dy^2}{\Delta_y^2} \Big.\nonumber\\
  &+  \bigg. \sum_{\nu=1}^\ell \sum_{p=0}^{m_\nu-1}\frac{(2x)^p e^{2x q_\nu}}{p!} \left(\bl{d\tau_\nu}^t\,\bl{P_{m_\nu,p-1}} \,\bl{d\tau_\nu} \right) \bigg] \,,
\end{align}
where the matrices $\bl{P_{m,p}}$ have been defined in Eq. (\ref{Pnk}). The functions $S_x$, $S_y$, $\Delta_x$ and $\Delta_y$ should be the ones found in the previous section while integrating Einstein's vacuum equation. In particular, note that the above line element is not diagonal, so that some of the Killing vector fields $\partial_{\tau_i}$ are not orthogonal to a family of hyper-surfaces.

Thus, now we have a recipe to explicitly construct new solutions for Einstein's vacuum equation in arbitrary dimensions. For each value of $n$ we can have solutions with different algebraic structures depending on the size of the blocks of the canonical form of $\bl{Q_0}$. For instance, when $n=4$  we have five possibilities: (I) the line element is diagonal, namely the canonical form of $\bl{Q_0}$ is the direct sum of one-dimensional blocks; (II) the canonical form of $\bl{Q_0}$ is the direct sum of two $2\times2$ blocks; (III) the canonical form of $\bl{Q_0}$ is the direct sum of a $2\times2$ block plus two $1\times1$ blocks; (IV) the canonical form of $\bl{Q_0}$ is the direct sum of a $3\times3$ block plus one $1\times1$ block; and (V) $\bl{Q_0}$ cannot be broken in smaller blocks by a similarity transformation, it is a single $4\times4$ block. In general, for a given $n$, the number of different algebraic structures for the line element  is the number of partitions of the integer $n$. In the next section we shall explore the possibilities for $n=2$ and $n=3$ in full detail.

%%%%%%%%%%%%%%%%%%%%%%%%%%%%%%%%%%%%%%%%%%%%%%%%%%%%%%%%%%%%%%%%%%%%%%%%%%%%%%%%%%
%%%%%%%%%%%%%%%%%%%%%%%%%%%%%%%%%%%%%%%%%%%%%%%%%%%%%%%%%%%%%%%%%%%%%%%%%%%%%%%%%%
%%%%%%%%%%%%%%%%%%%%%%%%%%%%%%%%%%%%%%%%%%%%%%%%%%%%%%%%%%%%%%%%%%%%%%%%%%%%%%%%%%
%%%%%%%%%%%%%%%%%%%%%%%%%%%%%%%%%%%%%%%%%%%%%%%%%%%%%%%%%%%%%%%%%%%%%%%%%%%%%%%%%%
%%%%%%%%%%%%%%%%%%%%%%%%%%%%%%%%%%%%%%%%%%%%%%%%%%%%%%%%%%%%%%%%%%%%%%%%%%%%%%%%%%
%%%%%%%%%%%%%%%%%%%%%%%%%%%%%%%%%%%%%%%%%%%%%%%%%%%%%%%%%%%%%%%%%%%%%%%%%%%%%%%%%%
%%%%%%%%%%%%%%%%%%%%%%%%%%%%%%%%%%%%%%%%%%%%%%%%%%%%%%%%%%%%%%%%%%%%%%%%%%%%%%%%%%
%%%%%%%%%%%%%%%%%%%%%%%%%%%%%%%%%%%%%%%%%%%%%%%%%%%%%%%%%%%%%%%%%%%%%%%%%%%%%%%%%%
%%%%%%%%%%%%%%%%%%%%%%%%%%%%%%%%%%%%%%%%%%%%%%%%%%%%%%%%%%%%%%%%%%%%%%%%%%%%%%%%%%
%%%%%%%%%%%%%%%%%%%%%%%%%%%%%%%%%%%%%%%%%%%%%%%%%%%%%%%%%%%%%%%%%%%%%%%%%%%%%%%%%%
%%%%%%%%%%%%%%%%%%%%%%%%%%%%%%%%%%%%%%%%%%%%%%%%%%%%%%%%%%%%%%%%%%%%%%%%%%%%%%%%%%
%%%%%%%%%%%%%%%%%%%%%%%%%%%%%%%%%%%%%%%%%%%%%%%%%%%%%%%%%%%%%%%%%%%%%%%%%%%%%%%%%%
%%%%%%%%%%%%%%%%%%%%%%%%%%%%%%%%%%%%%%%%%%%%%%%%%%%%%%%%%%%%%%%%%%%%%%%%%%%%%%%%%%

\section{Examples}\label{Sec.Examples}

In this section we work out explicit examples of the solutions found above, for both cases regarding whether the matrix $\bl{Q_0}$ is diagonalizable or not. Particularly, in the former case all the solutions turn out to be higher-dimensional generalizations of the Kasner metric. On the other hand, when $\bl{Q_0}$ is nondiagonalizable new solutions are attained. As stressed at the introduction, the special case $n=2$ of our solutions, namely when spaces are four-dimensional, has already been addressed previously in Ref. \cite{GabrielBatista2}. Thus, as we will check below, all solutions for $n=2$ should coincide with the ones of Ref. \cite{GabrielBatista2}.

\subsection{Diagonalizable Case}

In this subsection we shall deal with the case in which the complex symmetric matrix $\bl{Q_0}$ can be diagonalized. During the integration process of Einstein's equation, we had to consider separately two possibilities: (A) when $S_y'=0$; and (B) when $S_x'=0$ and $S_y'\neq0$. Likewise, here we shall treat these two possibilities separately.

\subsubsection{Case (A), $S_y'=0$}

We begin by analyzing the solution for the case $S_y'=0$, considering $\bl{Q_0}$ to be a diagonalizable matrix with eigenvalues given by $q_i$, for $i=1,\hdots,n$. In this case, the solution is given by the metric \eqref{LineElement5} with $S_y=0$, $\Delta_y=1$, while functions  $\Delta_x$ and $S_x$ are given respectively by \eqref{Deltax1} and  \eqref{Sx}. In addition, the constants $a_1$ and $a_2$ are related to each other by means of \eqref{c2}. In terms of the eigenvalues $q_i$, these constants are given by:
\begin{equation}\label{a0c2qi}
a_1=\sum_{i=1}^{n}q_i \quad \text{and} \quad a_2=(n+1)\sum_{i=1}^{n}q_i^2 \,,
\end{equation}
which is a consequence of Eq. (\ref{a0c2}). In particular, in terms of \eqref{a0c2qi}, relation \eqref{c2} becomes
\begin{equation}\label{sumqi}
\sum_{i=1}^{n}q_i^2=\frac{1}{(n+1)}\left(\sum_{i=1}^{n}q_i\right)^2+\frac{n s_1^{n+1}\Lambda}{c_1^2} \,.
\end{equation}

Now, defining constants $p_i$, for $i$ running from $1$ to $n$, and $p_{n+1}$ by
\begin{align}\label{pis}
p_i&=\frac{[(n+1)q_i-a_1]c_1}{(n+1)^{3/2}\sqrt{s_1^{n+1}\Lambda}}+\frac{1}{n+1}\,,\nonumber\\
p_{n+1}&=\frac{-a_1 c_1}{(n+1)^{3/2}\sqrt{s_1^{n+1}\Lambda}}+\frac{1}{n+1}\,,
\end{align}
and performing the coordinate transformation $(x,\tau_i,y)\rightarrow (r,\tilde{\tau}_i,\tilde{\tau}_{n+1})$ defined by
\begin{align*}
x&=\frac{c_1\log\left[\tan\big(\sqrt{(n+1)\Lambda}\,r/2\big)\right]}{\sqrt{(n+1)s_1^{n+1}\Lambda}}+x_0\,,\\
\tau_i&=\frac{2^{(p_i-\frac{1}{n+1})}e^{x_0(\frac{a_1}{n+1}-q_i)}}{s_1^{1/2}\big(\sqrt{(n+1)\Lambda}\,\big)^{p_i}}\tilde{\tau}_i\,,\\
y&=\frac{2^{(p_{n+1}-\frac{1}{n+1})}e^{\frac{a_1 x_0}{n+1}}}{s_1^{1/2}\big(\sqrt{(n+1)\Lambda}\,\big)^{p_{n+1}}}\tilde{\tau}_{n+1}\,,
\end{align*}
the solution becomes
\begin{equation}\label{KasnerND}
ds^2=dr^2+L_r^{2/(n+1)}\sum_{\alpha=1}^{n+1} e^{2\left(p_\alpha-\frac{1}{n+1}\right)N_r}(d\tilde{\tau}_\alpha)^2\,,
\end{equation}
where
\begin{align}\label{LrNr}
L_r&=\frac{\sin\big(\sqrt{(n+1)\Lambda}\,r\big)}{\sqrt{(n+1)\Lambda}} \,,\nonumber\\
N_r&=\log\Bigg[\frac{2 \tan\big(\frac{1}{2}\sqrt{(n+1)\Lambda}\,r\big)}{\sqrt{(n+1)\Lambda}} \Bigg]\,.
\end{align}
In this case, the parameters $p_\alpha$ obey the following relations:
\begin{equation}\label{pis2}
\sum_{\alpha=1}^{n+1}p_\alpha=1 \quad \text{and} \quad \sum_{\alpha=1}^{n+1}(p_\alpha)^2=1\,,
\end{equation}
the first relation being a direct consequence of definition \eqref{pis}, while the second one is obtained through Eq. \eqref{sumqi}.
Notice that, by setting $n=2$, namely for a four-dimensional space, the later solution becomes the Kasner metric generalized to contain a nonzero cosmological constant \cite{Stephani}.  Indeed, the metric \eqref{KasnerND}, along with \eqref{LrNr} and \eqref{pis2}, is the natural generalization of Kasner metric with cosmological constant for any number of dimensions. In particular, evaluating the limit $\Lambda\rightarrow 0$ we are lead to
\begin{equation}\label{Kasnernd}
ds^2=dr^2+\sum_{\alpha=1}^{n+1} r^{2p_\alpha}d\tilde{\tau}_\alpha^2\,,
\end{equation}
with the parameters $p_\alpha$ obeying \eqref{pis2}. This Ricci-flat metric is the $(n+2)$-dimensional version of the Kasner metric \cite{KasnerHighD}. Although we have not been able to find the full higher-dimensional solution \eqref{KasnerND}, with nonzero cosmological constant, in the literature, this can be seen as a simple generalization of the higher-dimensional Kasner metric found in Refs. \cite{KasnerHighD,KasnerHighD2}.

Similarly, the case where both the functions $S_x$ and $S_y$ are constant is given by the line element \eqref{LineElement5} with $S_x=s_1$, $S_y=0$, $\Delta_x=c_1 e^{-a_1 x}$ and $\Delta_y=1$. In this case, the eigenvalues of $\bl{Q_0}$ are constrained by the following expression:
\begin{equation}\label{pis3}
\bigg( \sum_{i=1}^{n} q_i \bigg)^2=\sum_{i=1}^{n}q_i^2\,.
\end{equation}
Then, defining constants $p_i=q_i/a_1$ and performing the coordinate transformation
\begin{align*}
\tau_i&=s_1^{(p_i-1)/2}(a_1 c_1)^{-p_i}\tilde{\tau}_i\,,\\
x&=a_1^{-1} \log \big[(a_1 c_1 r)/s_1^{1/2}\big]\,,\\
y&=s_1^{-1/2}\tilde{\tau}_{n+1}\,,
\end{align*}
the present solution reduces to \eqref{Kasnernd} with $p_{n+1}=0$, while the relation \eqref{pis3} boils down to \eqref{pis2}, again with $p_{n+1}=0$. Therefore, this solution  is a particular case of the $(n+2)$-dimensional Ricci-flat Kasner metric. It is worth mentioning that for the choice $n=2$ such solution degenerates to the four-dimensional flat space, in accordance with Ref. \cite{GabrielBatista2}.

\subsubsection{Case (B), $S_x'=0$ and $S_y'\neq0$}

Now we carry out the analysis of the case $S_x'=0$ and $S_y'\neq0$, with $\bl{Q_0}$ being a diagonalizable matrix with eigenvalues $q_i$, for $i=1,\hdots,n$. The solution for this case is given by the line element \eqref{LineElement4} with $\bl{Q_0}=\textrm{diag}(q_1,\cdots,q_n)$. Moreover, the following constraints must hold:
\begin{equation}\label{Constraint1}
  a_1 = \sum_{i=1}^n q_i \;\textrm{ and }\; \sum_{i=1}^n q_i^2 = \left( \sum_{i=1}^n q_i\right)^2 \,.
\end{equation}
In this solution, the function $S_y$ can be chosen to be any nonconstant function, as stressed out in the last paragraph of subsection \ref{SubsectionSx}. Therefore, for convenience, let us choose $S_y$ to be given by $S_y=-(n+1)/(\Lambda y^2)$, so that $\Delta_y=1$. Thus, performing the coordinate transformation  $(x,\tau_i,y)\rightarrow (r,\tilde{\tau}_i,\tilde{\tau}_{n+1})$
\begin{align*}
x&=a_1^{-1}\log(a_1  r)\,,\\
\tau_i&=(a_1 )^{-p_i}\tilde{\tau}_i\,,\\
y&= \tilde{\tau}_{n+1}\,,
\end{align*}
where we have defined $p_i=q_i/a_1$, it follows that the line element is given by
\begin{equation}\label{ExampleB}
ds^2 =-\frac{n+1}{\Lambda \tilde{\tau}_{n+1}^2} \Bigg[\,dr^2+\sum_{i=1}^{n} r^{2p_i}d\tilde{\tau}_i^2 + d\tilde{\tau}_{n+1}^2 \Bigg]\,.
\end{equation}
The constraints (\ref{Constraint1}) are now written in terms of $p_i$ as
\begin{equation*}
\sum_{i=1}^{n} p_i=\sum_{i=1}^{n} p_i^2=1\,.
\end{equation*}
This solution is conformal to the Kasner metric \eqref{Kasnernd} with $p_{n+1}=0$. As mentioned above, this particular case of Kasner metric reduces to the four-dimensional flat space when we set $n=2$ and, hence, solution (\ref{ExampleB}) becomes a four-dimensional maximally symmetric space when $n=2$, which is in perfect accordance with Ref. \cite{GabrielBatista2}.

Since we have defined $p_i=q_i/a_1$, the special case $a_1=0$ must be handled separately. Doing so, we obtain that
\begin{equation*}
\sum_{i=1}^{n} q_i=\sum_{i=1}^{n} q_i^2=0\,,
\end{equation*}
and the solution reduces to a line element already presented in the literature, see Ref. \cite{KasnerHD4}. Such spaces descend from Kasner spacetimes and their Kasner parameters $q_i$ are either all zero or some of them must be complex, in which case the spacetime may admit closed time-like curves \cite{KasnerHD4}.

%%%%%%%%%%%%%%%%%%%%%%%%%%%%%%%%%%%%%%%%%%%%%%%%%%%%%%%%%%%%%%%%%%%%%%%%%%%%%%%%%%%%%%%%%%%%%%%%%%%%%%%%%%%%%
%%%%%%%%%%%%%%%%%%%%%%%%%%%%%%%%%%%%%%%%%%%%%%%%%%%%%%%%%%%%%%%%%%%%%%%%%%%%%%%%%%%%%%%%%%%%%%%%%%%%%%%%%%%%%
%%%%%%%%%%%%%%%%%%%%%%%%%%%%%%%%%%%%%%%%%%%%%%%%%%%%%%%%%%%%%%%%%%%%%%%%%%%%%%%%%%%%%%%%%%%%%%%%%%%%%%%%%%%%%
%%%%%%%%%%%%%%%%%%%%%%%%%%%%%%%%%%%%%%%%%%%%%%%%%%%%%%%%%%%%%%%%%%%%%%%%%%%%%%%%%%%%%%%%%%%%%%%%%%%%%%%%%%%%%
%%%%%%%%%%%%%%%%%%%%%%%%%%%%%%%%%%%%%%%%%%%%%%%%%%%%%%%%%%%%%%%%%%%%%%%%%%%%%%%%%%%%%%%%%%%%%%%%%%%%%%%%%%%%%
%%%%%%%%%%%%%%%%%%%%%%%%%%%%%%%%%%%%%%%%%%%%%%%%%%%%%%%%%%%%%%%%%%%%%%%%%%%%%%%%%%%%%%%%%%%%%%%%%%%%%%%%%%%%%
%%%%%%%%%%%%%%%%%%%%%%%%%%%%%%%%%%%%%%%%%%%%%%%%%%%%%%%%%%%%%%%%%%%%%%%%%%%%%%%%%%%%%%%%%%%%%%%%%%%%%%%%%%%%%
%%%%%%%%%%%%%%%%%%%%%%%%%%%%%%%%%%%%%%%%%%%%%%%%%%%%%%%%%%%%%%%%%%%%%%%%%%%%%%%%%%%%%%%%%%%%%%%%%%%%%%%%%%%%%

\subsection{Nondiagonalizable Case}

Now, we shall deal with the most interesting case, namely when $\bl{Q_0}$ cannot be diagonalized. For $n\geq 3$, this case leads to solutions that, as far as the authors know, have not been described in the literature yet. But, first, let us start considering the case $n=2$ and showing that the solution obtained here coincides with the one of Ref. \cite{GabrielBatista2}. Then, we shall consider the case $n=3$.

\subsubsection{Case $n=2$}

In this subsection we provide a few examples of metrics built from a nondiagonalizable $\bl{Q_0}$ when $n=2$. Using the construction exhibited in section \ref{Sec.NonDiagTheory}, it follows that in such a case $\bl{Q_0}$ admits just one eigenvalue, here denoted by $q$, and there exists an orthogonal complex matrix $\bl{M}$ such that $\bl{M}^{t} \bl{Q_0 M} =\bl{Q_1}$ assumes the following canonical form:
\begin{equation*}
  \bl{Q_1}= q \bl{I_2}+\bl{P_{2,0}} =  q\left[ \begin{array}{cc}
1& 0 \\
0 & 1 \\
\end{array}
\right] + \frac{1}{2}\left[ \begin{array}{cc}
i & 1 \\
1 & -i \\
\end{array}
\right]\,.
\end{equation*}
Thus, in accordance with Eq. (\ref{ExpQ}), the exponential of $2x\bl{Q_1}$ reads:
\begin{align*}
  e^{2x\bl{Q_1}} &=  e^{2x q} \sum_{p=0}^{1}\frac{(2x)^p}{p!}\bl{P_{2,p-1}} \\
  & = e^{2x q}
  \left[ \begin{array}{cc}
1+i x & x \\
x & 1-i x \\
\end{array}
\right]\,,
\end{align*}
where it has been used the fact that $\bl{P_{2,-1}}$ is simply the identity, as can be checked from \eqref{Pnk}.
Thus, the part of the line element involving the exponential is given by
\begin{align*}
  \bl{d\tau_1}^t &\,e^{2x\bl{Q_1}}\, \bl{d\tau_1}    \\
&=e^{2x q}
  \left[ \begin{array}{cc}
d\tau_1 & d\tau_2 \\
\end{array}
\right]
  \left[ \begin{array}{cc}
1+i x & x \\
x & 1-i x \\
\end{array}
\right]
  \left[ \begin{array}{c}
d\tau_1 \\
d\tau_2\\
\end{array}
\right]\\
&=e^{2x q}\left[(1+ix)d\tau_1^2+(1-ix)d\tau_2^2+2x d\tau_1 d\tau_2\right]\,.
\end{align*}
Notice that this result, along with $tr(\bl{Q_0})=tr(\bl{Q_1})=2q$ and $tr(\bl{Q_0}^2)=tr(\bl{Q_1}^2)=2q^2$, is valid for all the cases regarding the constancy of the functions $S_x$ and $S_y$.

Thus, for instance, for the case (A), when $S_y$ is a constant, the solution is given by the line element (\ref{LineElement3}), which now reads
\begin{align*}
  ds^2 = S_x\bigg\{   e^{2x q}&\left[(1+ix)d\tau_1^2+(1-ix)d\tau_2^2+2x d\tau_1 d\tau_2\right] \Big.\nonumber\\
  & +\frac{dx^2}{\Delta_x^2} + dy^2 \bigg. \bigg\} \,,
\end{align*}
with functions $S_x$ and $\Delta_x$ given by
\begin{align*}
S_x&=s_1\big\{ e^{2qx} \cosh[q(x-x_0)]  \big\}^{-2/3}\,,\\
\Delta_x&=\frac{\sqrt{3s_1^3\Lambda}}{q}\, e^{-2q x}S_x^{-1}\,.
\end{align*}
Then, performing the coordinate transformation $(\tau_1,\tau_2,x,y)\rightarrow (\phi, t, r, \theta)$ below:
\begin{align*}
\tau_1&=\frac{e^{3i\pi/4-q x_0/3}}{2^{11/6}s_1^{1/2}q^{1/2}}[\phi-(2(x_0+i)q-\log\Lambda)t]\,,\\
\tau_2&=\frac{e^{i\pi/4-q x_0/3}}{2^{11/6}s_1^{1/2}q^{1/2}}[\phi-(2(x_0-i)q-\log\Lambda)t]\,,\\
x&=\frac{1}{2q}(r-\log\Lambda)+x_0\,,\\
y&=\frac{e^{2qx_0/3}\theta}{2^{1/3}s_1^{1/2}\Lambda^{1/2}}\,,
\end{align*}
we obtain the following line element:
\begin{equation}\label{metric23}
ds^2=\frac{e^{-r} dr^2}{3(1+\Lambda e^{-r})^2}+\frac{e^{-r}d\theta^2-dt(r dt+d\phi)}{(1+ \Lambda e^{-r} )^{2/3}}\,.
\end{equation}
This solution is a Kundt spacetime of Petrov type II, which has recently been obtained in Ref. \cite{GabrielBatista2}, and, therefore, represents no novelty. Nevertheless, the current work helps to shed light on the origin of such solution, inasmuch as we have just proved that (\ref{metric23}) is nothing more than the simplest member of an arbitrary-dimensional class of solutions of Einstein's vacuum equation depending on the exponential of a non-diagonalizable symmetric matrix.

For the case (B), where $S_x'=0$ and $S_y\neq 0$, a similar treatment provides that the solution is a maximally symmetric space, as already anticipated in Ref. \cite{GabrielBatista2}. In such case, we notice that even though the part $\bl{d\tilde{\sigma}}^t e^{2x\bl{Q_0}} \bl{d\tilde{\sigma}}$  of the line element cannot be diagonalized using cyclic coordinates, the metric as a whole can.

\subsubsection{Case $n=3$}

As a second example, let us consider $n=3$. In this case we have three different algebraic types for the canonical form of the matrix $\bl{Q_0}$: (I) it can be diagonal; (II) it can be the direct sum of a $2\times 2$ plus a $1\times 1$ blocks; (III) it can be a single  $3\times 3$ block. The case (I) has already been considered above for arbitrary $n$, whereas the case (II) can be easily tackled by adding the term $e^{2x q_2}d\tau_3^2$ to the line element obtained in the previous subsection (case $n=2$), in accordance with Eq. (\ref{Nondiagmetric}), which yields
\begin{align*}
  ds^2 &= (S_x + S_y)\bigg\{ \frac{dx^2}{\Delta_x^2} + \frac{dy^2}{\Delta_y^2} + e^{2x q_2}d\tau_3^2\Big.\nonumber\\
  &+ e^{2x q_1}\left[(1+ix)d\tau_1^2+(1-ix)d\tau_2^2+2x d\tau_1 d\tau_2\right] \bigg\} \,.
\end{align*}
In order to eliminate the complex character of this metric we can perform the change of coordinates $(\tau_1,\tau_2)\rightarrow (\tilde{\tau}_1,\tilde{\tau}_2)$ defined by
\begin{align*}
\tau_1=\frac{e^{3i\pi/4}}{2}(\tilde{\tau}_1+i\tilde{\tau}_2) \quad \text{and} \quad \tau_2=\frac{e^{5 i\pi/4}}{2}(\tilde{\tau}_1-i\tilde{\tau}_2)\,,
\end{align*}
so that the line element reads
\begin{align}
  ds^2 = (S_x + S_y)\bigg[&e^{2x q_1} d\tilde{\tau}_1(x d\tilde{\tau}_1+d\tilde{\tau}_2) + e^{2x q_2}d\tau_3^2 \nonumber\\
  &+\frac{dx^2}{\Delta_x^2} + \frac{dy^2}{\Delta_y^2} \bigg] \,. \label{2+1}
\end{align}
As usual, the functions $\Delta_x$,  $\Delta_y$, $S_x$ and $S_y$ should be the ones obtained through the integration process of Einstein's equation, see Sec. \ref{Sec.Integration}. Moreover, $q_1$ and $q_2$ are related to each other through Eq. (\ref{c2}) or (\ref{c2c1}).
In spite of the fact that this solution has not been described in the literature yet (as far as the authors know), it can be seen as a simple extension of the solution for the case $n=2$, which have been recently found in Ref. \cite{GabrielBatista2}. Differently, the possibility (III) of the case $n=3$, namely when the canonical form of $\bl{Q_0}$ is a single $3\times 3$ block, leads to a solution that has a new algebraic structure and, therefore, cannot be seen as a simple generalization of the former examples. Now, let us consider such possibility.

When $\bl{M}^t \bl{Q_0 M}=\bl{Q_1}$ is a single $3\times 3$ block with eigenvalue $q$, we have that
\begin{align}
  e^{2x\bl{Q_1}} &=  e^{2x q} \sum_{p=0}^{2}\frac{(2x)^p}{p!}\bl{P_{3,p-1}}  \nonumber\\
  &=e^{2x q}(\bl{P_{3,-1}}+2x\bl{P_{3,0}}+2x^2\bl{P_{3,1}})\,,	 \label{eQ1-n3}
\end{align}
where $\bl{P_{3,-1}}$ is the identity matrix, whereas $\bl{P_{3,0}}$ and $\bl{P_{3,1}}$ are given by
\begin{align*}
\bl{P_{3,0}}= \frac{1}{2}\!
\left[\! \begin{array}{ccc}
0 & 1+i & 0 \\
1+i & 0 & 1-i \\
0 & 1-i & 0 \\
\end{array}
\!\!\right]\!,\,
\bl{P_{3,1}}= \frac{1}{2}\!
\left[ \begin{array}{ccc}
i & 0 & 1 \\
0 & 0 & 0 \\
1 & 0 & -i \\
\end{array}
\right]\!.
\end{align*}
Thus, inserting these into Eq. (\ref{eQ1-n3}) we are led to
\begin{align*}
  \bl{d\tau_1^t}  \,e^{2x\bl{Q_1}}\,  \bl{d\tau_1} = & \, e^{2x q}\bigg[
(1+ix^2)d\tau_1^2+d\tau_2^2  \\
&+(1-ix^2)d\tau_3^2 +2(1+i)x d\tau_1 d\tau_2 \\
&+2x^2 d\tau_1 d\tau_3+2(1-i)x d\tau_2 d\tau_3
\bigg]\,.
\end{align*}
In order to get rid of the complex dependence in the line element one can perform the coordinate transformation $(x,\tau_3)\rightarrow (\tilde{x},\tilde{\tau}_3)$ defined by
\begin{equation*}\label{coordtranf}
x=-(1-i)\tilde{x}\quad \text{and} \quad \tau_3=i \tilde{\tau}_3\,,
\end{equation*}
along with a redefinition of the constant parameter $q$, $q \rightarrow \tilde{q} =-(1-i)q$, which easily leads to
\begin{align*}
& \bl{d\tau_1^t}  \,e^{2x\bl{Q_1}}\,  \bl{d\tau_1} =e^{2\tilde{x} \tilde{q}}\bigg[
(1+2\tilde{x}^2)d\tau_1^2+d\tau_2^2 \\
&-(1-2\tilde{x}^2)d\tilde{\tau}_3^2 \,-4\tilde{x}d\tau_1 (d\tau_2-\tilde{x}d\tilde{\tau}_3)-4\tilde{x} d\tau_2 d\tilde{\tau}_3\bigg]\,. \\
\end{align*}

Since we have incorporated a complex factor into the coordinate $x$, it may seem that we have spoiled the expressions for $S_x$ and $\Delta_x$ with complex numbers. But this is not the case. Indeed, in the expressions for these functions, see Eqs. (\ref{Deltax1}), (\ref{Sx}), and (\ref{DxDyB}), the coordinate $x$ appears in the combinations $a_1 x$ and $\sqrt{a_2 - a_1^2} x$. But, since
\begin{equation}\label{a1a2Example}
  \left.
    \begin{array}{ll}
      a_1 &= \textrm{tr}\left(\bl{Q_0}\right) = \textrm{tr}\left(\bl{Q_1}\right) = 3q \,, \textrm{ and} \\
       a_2 &= 4\,\textrm{tr}\left(\bl{Q_0}^2\right) = 4\,\textrm{tr}\left(\bl{Q_1}^2\right) = 12 q^2\,,
       \end{array}
  \right.
\end{equation}
it follows that
\begin{equation*}
  a_1 \, x =  3 \tilde{q} \tilde{x}  \, , \;\;\textrm{ and }\;\; \sqrt{a_2 - a_1^2}\,x =   \sqrt{3} \tilde{q} \tilde{x} \,,
\end{equation*}
so that, at the end of the day, no complex factor shows up. Regarding the complex factor coming from the term $dx^2$, it can be trivially absorbed into the integration constant $c_1$.

Thus, using these results along with Eqs. (\ref{Deltax1}), (\ref{Sx}), (\ref{c2}), and (\ref{LineElement3}), it follows that, for the case (A) of the integration process, when $S_y'=0$, the final solution reads
\begin{widetext}
\begin{equation*}
  ds^2 =  S_{\tilde{x}}\left\{ e^{2\tilde{x} \tilde{q}}\bigg[ (1+2\tilde{x}^2)d\tau_1^2+d\tau_2^2  -(1-2\tilde{x}^2)d\tilde{\tau}_3^2 \,-4\tilde{x}d\tau_1 (d\tau_2-\tilde{x}d\tilde{\tau}_3)-4\tilde{x} d\tau_2 d\tilde{\tau}_3\bigg] +
   \frac{\tilde{q}^2 e^{6 \tilde{q}\, \tilde{x}}}{4 s_1^4 \Lambda}\,S_{\tilde{x}}^3 d\tilde{x}^2+ dy^2 \right\}\,,
\end{equation*}
\end{widetext}
where the function $S_{\tilde{x}}$ is defined by
\begin{equation*}
  S_{\tilde{x}} = s_1\,\left\{e^{3 \tilde{q} \tilde{x}}\cosh\!\left[\, \tilde{q} (\tilde{x}-\tilde{x}_0) \,\right]\right\}^{-1/2}\,.
\end{equation*}
The constants $s_1$ and  $\tilde{x}_0$ are arbitrary, but they are both nonphysical, as we can set them to $1$ and $0$ respectively by means of a coordinate transformation. Actually, in all solutions presented in this article, for arbitrary $n$, we can set $s_1=1$ and $x_0=0$ without changing the geometry of the spacetime.  Likewise, we can also set $\tilde{q}=1$ in the latter line element, which just amounts to a coordinate transformation. Thus, the only important parameter in this metric is the cosmological constant, $\Lambda$. As far as the authors now, this solution have not been described before in the literature.

%The constants $s_1$ and  $\tilde{x}_0$ are arbitrary, but they are both nonphysical, as we can set them to $1$ and $0$ respectively by means of a %coordinate transformation. As far as the authors now, this solution have not been described before in the literature.

Concerning the case (B) for $n=3$, when $S_x'=0$ and  $S_y'\neq 0$, we should have $a_2 = 4 a_1^2$, as a consequence of Eq. (\ref{c2c1}). However, generally this restriction is compatible with Eq. (\ref{a1a2Example}) only if we set $\tilde{q}=0$. Actually, this conclusion of vanishing eigenvalue is valid for arbitrary values of $n$, whenever the canonical form of $\bl{Q_0}$ is a single block. Thus, for the case $n=3$ with $S_x'=0$ we eventually arrive at the following solution:
\begin{widetext}
\begin{equation*}
  ds^2 =  - \frac{4}{\Lambda\,y^2}\left\{  \bigg[ (1+2\tilde{x}^2)d\tau_1^2+d\tau_2^2  -(1-2\tilde{x}^2)d\tilde{\tau}_3^2 \,-4\tilde{x}d\tau_1 (d\tau_2-\tilde{x}d\tilde{\tau}_3)-4\tilde{x} d\tau_2 d\tilde{\tau}_3\bigg] + \tilde{c}_1\, d\tilde{x}^2+ dy^2 \right\}\,.
\end{equation*}
\end{widetext}

\section{Conclusions and Perspectives}\label{Sec.Conc}

Starting with the class of $(n+2)$-dimensional spaces of Eq. (\ref{LineElement1}), we fully integrate Einstein's vacuum equation with a cosmological constant. The solutions found in Sec. \ref{Sec.Integration} turn out to depend on the exponential of an arbitrary  constant symmetric matrix $\bl{Q_0}$. By the use of suitable coordinate transformations, this exponential can be explicitly computed and the attained line element can have different algebraic structures depending on the canonical form of $\bl{Q_0}$. For example, in the simplest case, when $\bl{Q_0}$ is diagonalizable, the solutions are higher-dimensional generalizations of Kasner metric, which represent homogeneous but non-isotropic spacetimes used in cosmological models. However, when $\bl{Q_0}$ cannot be diagonalized more interesting solutions are obtained. As far as the authors know, such solutions have not been described before in the literature.

In the general solution obtained here, the part of the line element associated with the cyclic coordinates $\tau_i$ is generally composed by the sum of smaller blocks, see Eq. (\ref{Nondiagmetric}). Thus, once we have computed the structure of the line element in the cases in which the canonical forms of $\bl{Q_0}$  are single $n\times n$ blocks, for different values of $n$, we can compose these solution in order to generate Einstein spaces of higher dimensions. For instance, in Eq. (\ref{2+1}) we have merged a $2\times 2$ block with a $1\times 1$ block in order to create a solution with $n=3$.

In the present work we have focused exclusively on the integration of Einstein's equation and simplification of the algebraic structure of the solutions. The physical aspects, as well as the geometrical properties of these solutions, have not been addressed. In the future we intend to fill these gaps in order to attain a full comprehension of such spaces.

\vspace{0.5cm}

\begin{acknowledgments}
C. B. would like to thank Conselho Nacional de Desenvolvimento Cient\'{\i}fico e Tecnol\'ogico (CNPq) for the partial financial support through the research productivity fellowship. Likewise,  C. B. thanks Universidade Federal de Pernambuco for the funding through Qualis A project.  G. L. A. thanks CNPq for the financial support.
\end{acknowledgments}

\end{document}